\documentclass[twocolumn]{article}
\usepackage{graphicx}
\usepackage{eqnarray}
\usepackage{amssymb}
\usepackage{anysize}
\usepackage{amsfonts,amssymb,latexsym,amsmath,amscd}
\marginsize{25mm}{20mm}{35mm}{40mm} \textheight=215mm

\begin{document}
    \title{\textbf{Electro-Optic Effect Explanation with Quantum Photonic Model}}
    \author{Hassan Kaatuzian$^{1}$\and AliAkbar Wahedy Zarch$^{1}$\footnote{Material presented in this
    paper is a part of AliAkbar Wahedy Zarch work on his
    thesis towards Ph.D degree. Dr. Hassan Kaatuzian is his supervisor on thesis, fatemahali@aut.ac.ir.}
    \and Ahmad Amjadi$^{2}$ \and Ahmad Ajdarzadeh Oskouei$^{3}$\\ \\$^{1}$\normalsize Photonics Research Laboratory (PRL), Electrical Engineering Department\\
    \normalsize AmirKabir University of Technology, Tehran, IRAN, hsnkato@aut.ac.ir \\
    $^{2}$\normalsize Physics Department, Sharif University of Technology, Tehran,
    IRAN, amjadi@sharif.edu.\\
    $^{3}$\normalsize Laboratory of Ultrafast Spectroscopy, Ecole Polytechnique Federale
    de Lausanne(EPFL),\\ \normalsize Switzerland, ahmad.ajdarzadeh@epfl.ch.}
    \date{}
    \maketitle

    \begin{abstract}
       In this paper, we have explained transverse electro-optic effect
       by quantum-photonic model (QPM). This model interpret this effect by
       photon-electron interaction in attosecond regime. We
       simulate applied electric field on molecule and crystal by
       Monte-Carlo method in time domain when a light beam is propagated through the waveguide. We show how the waveguide response to an optical
       signal with different wavelengths when a transverse electric field applied to the waveguide.
    \end{abstract}

    \vspace{2pc} \noindent{\it Keywords}: Quantum-photonic model(QPM),
        photon-electron interaction, attosecond regime, electro-optic (EO) effect, NPP.\\

    \section{\large Introduction}
        Electro-optic polymers are particularly interesting for new device
        design and high-speed operation \cite{1}-\cite{7}. Organic optical materials like MNA, NPP, MAP have a high figure
        of merit in optical properties in comparison with inorganic optical materials such as BBO, LiNbO$_{3}$, \cite{8}.
        2-methyl-4-nitroaniline (MNA) and N-(4-nitrophenyl)-L-prolinol (NPP) have the highest figure
        of merit between organic nonlinear optical materials \cite{9}. Thus they are used for electro-optic
        and nonlinear optic applications\cite{8}-\cite{21}.
        The most effective element in optical phenomena could be the refractive index.
        Because of the large electro-optic coefficient of organic material, a certain amount of
        refractive index (RI) change can be realized with lower driving voltage
        than in other EO materials($1.5 V/\mu m$ for NPP,\cite{10}).
        Ledoux et al. work on linear and nonlinear properties of NPP crystal. They first proposed a Sellmeier set for
        RI of NPP that is based on measured refractive indices\cite{11}. Banfi, Datta and co-workers design a more accurate
        Sellmeier set for RI of NPP for nonlinear optical studies \cite{13}-\cite{15}. Their data is based on
        classic and macroscopic measurements and approaches. Some physicians
        explain RI in molecular bases for typical material \cite{22}-\cite{25}. Some authors have calculated RI of real material
        by quantum mechanical approach,\cite{26}-\cite{30}. Recently it is done some calculations and simulations on
        RI of liquid crystal by some softwares based on quantum mechanics (QM) \cite{31}.
        In this paper, EO effect
        has been analyzed using QPM. We use QPM to explain linear optical
        phenomena in molecular scales. This model gives us a constitutive
        vision about phenomena and real materials \cite{32}-\cite{35}.
        This approach is based on four elements: 1-Quasi-classic principle for justification of optical phenomenon in
        molecular scales, 2- Knowledge of crystal network and its space shape, 3-
        Short range intramolecular and intermolecular forces, 4-
        Monte-Carlo time domain simulation. In this model we suppose a laser beam is a
        flow of photons while passing through single crystal film, interacts
        with delocalization $\pi$-electron system of organic molecule and delays
        the photon in every layer \cite{35}. By precise calculation of these
        retardation in every layer, we obtain RI in
        specific wavelength and explain EO effect too. We show that the phase retardation
        of input light with different wavelengths is distinctive, when it travels through the waveguide.
        The results obtained from this method are
        well agreed with experimental data.
        Our favorite organic molecule is
        NPP.

    \section{\large Crystal and Molecular Structure of our Favorite Organic Compound}

        Organic molecular units and conjugated polymer chains possessing  $\pi$-electron
        systems usually form as centrosymmetric structures  and thus, in the electric
        dipole approximation, do not show any linear electro-optic and nonlinear second order optical properties.
        The necessary acentric may be provided by first distorting the $\pi$-electron system
        by interaction with strong electron donor and acceptor groups \cite{36}.
        In NPP molecule, nitro group acts as an acceptor and the other main groups on the other side of benzene
        ring acts as a week donor (see fig. 1).
        \begin{figure}
            \centering
            \includegraphics[scale=0.25]{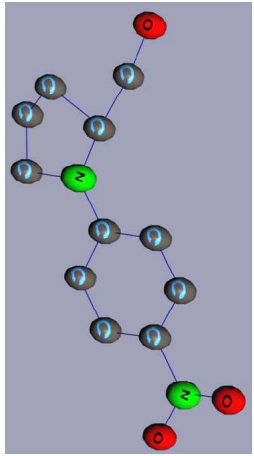}
            \caption{\footnotesize The molecular compound of NPP.}\label{Fig1}
        \end{figure}
        NPP $(C_{11}H_{14}N_{2}O_{3})$ (Fig.1) crystallizes in the solid
        state in an acentric monoclinic (with space group $P2_{1}$) structure and
        their parameters are:a=5.261$A^{\circ}$,b=14.908$A^{\circ}$,c=7.185$A^{\circ}$
        ,$\beta$=105.18$^{\circ}$ and in the wavelength range of 0.5 to 2$\mu$m is transparent. The most
        interesting property of NPP crystal
        is the proximity of the mean plane of molecule with the crystallographic
        plane (101); the angle between both of these planes being 11$^{\circ}$.
        Nitro group of one molecule in downward connects to Prolinol group in upper
        by hydrogen bonding. The angle between b orientation of crystal and
        N(1)-N(2) axis (charge-transfer axis) is equal to 58.6$^{\circ}$ \cite{8}. Fig. 2 shows the crystal packing of the NPP.
        \begin{figure}
            \centering
            \includegraphics[scale=0.22]{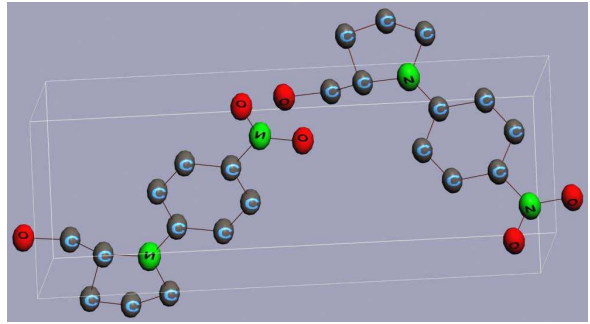}
            \caption{\footnotesize The crystal packing of NPP.}\label{Fig2}
        \end{figure}
        For accurate and valid simulation, these properties and angles have
        to be exerted. For benzene molecule, benzene ring is a circle (see fig.\ref{Fig3});
        \begin{figure*}
            \centering
            \includegraphics[scale=1.2]{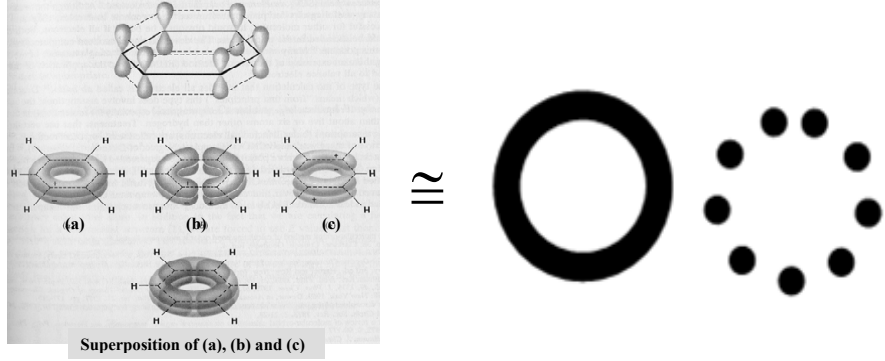}
            \caption{\footnotesize Electron cloud for Benzene molecule that obtained from Huckle theory \cite{37} and its
            approximation.}\label{Fig3}
        \end{figure*}
        Fig.\ref{Fig4} demonstrates
        \begin{figure}
            \centering
            \includegraphics[scale=0.35]{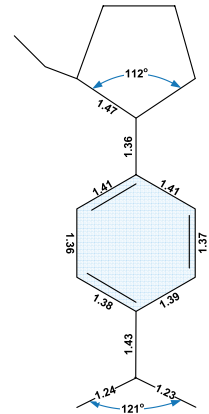}
            \caption{\footnotesize The bond lengths (in angstrom) and angles of NPP molecule.}\label{Fig4}
        \end{figure}
        the bond lengths and angles of the NPP molecule. As we see in this figure the bond lengths
        in benzene ring are not same. In our simulation, for similarity
        we consider an ellipse correspond to circle for electron cloud.
        We obtained $\varepsilon$=0.26 for the ellipse of NPP from simulation, fig.\ref{Fig5} shows the comparison
        of a circle and an ellipse with $\varepsilon$=0.26.
        \begin{figure}
            \centering
            \includegraphics[scale=0.6,width=3in,height=3in]{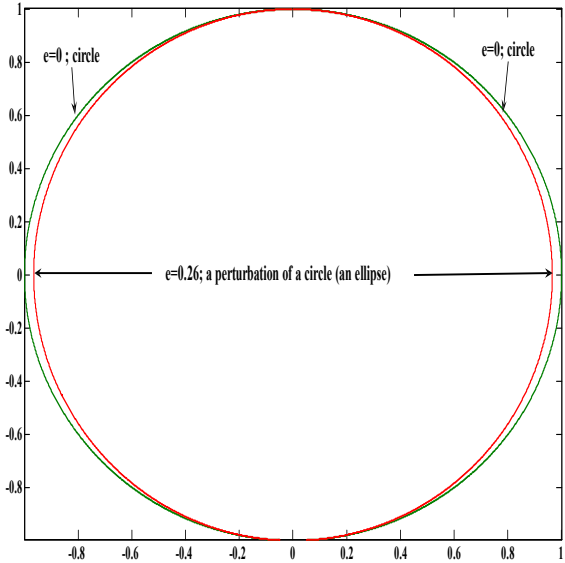}
            \caption{\footnotesize The comparison between circles for Benzene
            molecule electron cloud and NPP molecule electron cloud, (approximately). As we see the electron cloud
            of $\pi-$electron in NPP has distorted a little.}\label{Fig5}
        \end{figure}

    \section{\large Our Model for Electro-Optic Effect}

        For a biaxial crystal that $n_{x}\neq n_{y}\neq n_{z}$ the equation of index ellipsoid is
        \begin{equation}
            \label{eq1}
        \frac{x^{2}}{n_{x}^{2}}+\frac{y^{2}}{n_{y}^{2}}+\frac{z^{2}}{n_{z}^{2}}=1
        \end{equation}
        Assume that single crystal
        film lies in x-y plane and light propagation is in the
        direction; therefore in
        the presence of an electric field the equation of index
        ellipsoid by assuming crystal symmetry  will become:
        \begin{equation}
            \label{eq2}
            [(\frac{1}{n^{2}})_{_{x}}+r_{12}E_{y}]x^{2}+
            [(\frac{1}{n^{2}})_{_{y}}+r_{22}E_{y}]y^{2}+
            2r_{61}E_{x}xy=1
        \end{equation}
        for NPP that $E_{x}$ and $E_{y}$ are transverse electric field components,
        \cite{10}. For MNA crystal Eq.(2) will become:
        {\setlength\arraycolsep{2pt}
         \begin{eqnarray}
            \label{eq3}
            [(\frac{1}{n^{2}})_{_{x}}+r_{11}E_{x}]x^{2}+[(\frac{1}{n^{2}})_{_{y}}+r_{21}E_{x}]y^{2}
            \nonumber\\
            {}+[(\frac{1}{n^{2}})_{_{z}}+r_{31}E_{x}]z^{2}+2r_{51}E_{x}xz=1
            \nonumber\\
        \end{eqnarray}
        where $E_{x}$ is only transverse electric field component \cite{21}.
        With appropriate rotational transformation, these
        relations can been simplified. In NPP, $r_{12}$ and $r_{22}$ and
        in MNA, $r_{11}$ and $r_{21}$ is large coefficient. Therefore
        $n_{x}$ and $n_{y}$ for NPP simplified to
        $n_{x}-(1/2)n_{x}^{3}r_{12}E_{y}$ and
        $n_{y}-(1/2)n_{y}^{3}r_{22}E_{y}$ and for MNA $n_{x}-(1/2)n_{x}^{3}r_{11}E_{x}$
        and $n_{y}-(1/2)n_{y}^{3}r_{21}E_{x}$ respectively. The
        phase retardation $\Gamma$, with an applied electric field
        in a typical linear transverse electrooptic modulator will
        be obtained as follows, \cite{38}:
        \begin{equation}
            \label{eq4}
            \Gamma=\acute{\phi_{y}}-\acute{\phi_{x}}=\frac{\omega l}{c}\cdot[n_{y}-n_{x}-\frac{1}{2}(n_{y}^{3}r_{22}-n_{x}^{3}r_{12})E_{y}]
        \end{equation}
        for NPP and
        \begin{equation}
            \label{eq5}
            \Gamma=\acute{\phi_{y}}-\acute{\phi_{x}}=\frac{\omega l}{c}\cdot[n_{y}-n_{x}-\frac{1}{2}(n_{x}^{3}r_{11}-n_{y}^{3}r_{21})E_{x}]
        \end{equation}
        for MNA crystal. The total phase difference between two
        perpendicular polarization of light (in our example $E_{x}$ and
        $E_{y}$),is
        \begin{equation}
            \label{eq6}
            \Delta=\Delta_{0}+\delta
        \end{equation}
        where $\Delta_{0}$ is due to linear birefringence and
        $\delta$ is due to linear electro-optic effect. In this
        case $\delta$ is much smaller than $\Delta_{0}$.\\
        In sub-micron space scales and sub-femtoseconds time scales,
        the optical constants loses its stabilization and classical
        equations for linear and nonlinear optical phenomena are not useful \cite{45}.
        Now, we suggest a microscopic model for linear electro-optic phenomenon.
        If we have a monochromatic
        laser beam with frequency $\nu$ and intensity I, then we may attain average photon
        flux from relation \cite{39}:
        \begin{equation}
            \label{eq7}
            \phi=\frac{I}{h\nu}
        \end{equation}
        now if we assume thin single crystal film of NPP in $"b"$ direction of
        crystal (or $"z"$ axis) radiated by a He-Ne laser with: $\lambda$=633 nm,
        average power=10mw and beamwidth=20 microns, then from (\ref{eq4}) average photon
        flux is equal to $10^{22} photons/(s-cm^{2})$ that signifies in every second
        $10^{22}$ photons arrive to each centimeter square. Moreover from data of
        crystal in subsection $(A)$ in every 36.5$(A^{\scriptstyle{\circ^{2}}})$
        on z direction, one NPP molecule exists.
        Therefore in every second $36.5\times10^{6}$
        photons interact with any molecule or
        in other words in every 27ns, (with assuming only
        linear optic phenomenon exist and
        nonlinear optic phenomenon do not exist, approximately. Because laser watt is not much),
        one photon interacts with any NPP molecules.
        In each interaction between photon and electron
        in every layer of crystal,
        we suppose a delay time equal to $\tau_{i}$ ($i$th layer of the crystal).
        Total delay time for m layers in crystal region is equal to:
        \begin{displaymath}
            \sum_{i=1}^{m}\tau_{i}
        \end{displaymath}
        Consequently required time for photon transmission in L length
        of crystal is equal to $\tau$, achieved from relation:
        \begin{eqnarray}
            \label{eq8}
            \tau=\frac{L}{\frac{c_{0}}{n}}=\frac{nL}{c_{0}}=\frac{L}{c_{0}}+\sum_{i=1}^{m}\tau_{i}
        \end{eqnarray}
        where $c_{\scriptscriptstyle{0}}$ is velocity of light in vacuum.
        By using this relation, we can relate
        macroscopic quantity $n$ to microscopic quantity $\tau_{i}$.
        In biaxial crystals, $\tau$ and consequently $n$ depends on polarization direction
        of incident light. Because dipole-field interaction conclusion
        is different for any direction of molecule. If $\tau_{x}$
        would be a microscopic delay for interaction of $x$-polarization field with dipole (or charge transfer
        action) and $\tau_{y}$ would be a microscopic delay for interaction of $y$-polarization field with dipole then the
        final phase difference between these two fields (named phase
        retardation) will be:
        \begin{equation}
            \label{eq9}
            \Delta\phi=\omega\cdot(\tau_{x}-\tau_{y})=\omega\cdot\sum_{i}(\tau_{x_{i}}-\tau_{y_{i}})
        \end{equation}
        Of course this relation give $\Delta_{0}$ of (6). We justify
        $\delta$ from our model in later subsections.

        \section{\large Photon-Electron Interaction in Attosecond Regime}

        For analyzing the interaction of the electric field of the
        photons with $\pi$-electron system, the time-dependent
        Schroedinger equation has to be used:
        \begin{equation}
            \label{eq10}
            H(r,t).\Psi(r,t)=\imath\hbar\frac{\partial\Psi(r,t)}{\partial t}
        \end{equation}
        with the Hamilton operator H representing the total energy
        of the matter-light system and the wave function $\Psi$
        representing the quantum state of this system with all
        detailed spatial and temporal information of all particles in
        it. First, the stationary Schrodinger for matter without
        any external interaction is usually applied:
        \begin{equation}
            \label{eq11}
            H_{matter}(r).\varphi_{m}(r)=E_{m}.\varphi_{m}(r)
        \end{equation}
        The interaction of the photon field with $\pi$-electron
        system can be described by first-order perturbation
        theory. The Hamiltonian of (10) is split into the material
        steady-state Hamiltonian of (11) and the Hamiltonian of
        the interaction as a small disturbance:
        \begin{equation}
            \label{eq12}
            H(r,t)=H_{matter}+H_{interaction}(t)
        \end{equation}
        With this equation the temporal change of the coefficient
        describing the transitions of the the particle under the
        influence of the light can be calculated from:
        \begin{equation}
            \label{eq13}
            \frac{\partial}{\partial
            t}c_{p}(t)=-\frac{\imath}{\hbar}\sum_{m=1}^{\infty}
            [c_{m}(t).\int_{V}\psi_{p}^{*}H_{interaction}\psi_{m}dV]
        \end{equation}
        with the integration over the whole volume V of the
        wavefunctions. The probability of the population of state
        p is given by the square of $c_{p}$ and the transition
        probability $\omega_{p\leftarrow m}$ for the transition
        from state m to state p is given by:
        \begin{equation}
            \label{eq14}
            \omega_{p\leftarrow m}=\frac{\partial}{\partial
            t}|c_{p}(t)|^2\propto \mu_{p\leftarrow m}^2
        \end{equation}
        wich is proportional to the square of the transition
        dipole moment $\mu_{p\leftarrow m}$:
        \begin{equation}
            \label{eq15}
            \mu_{p\leftarrow m}=\int_{V}\varphi_{p}^{*}.H_{interaction}.\varphi_{m}dV
        \end{equation}
        The interaction operator is given for a one-electron
        system in the dipole approximation, assuming a radiation
        wavelength large compared to the dimension of the
        particle, by:
        \begin{equation}
            \label{eq16}
            H_{interaction}(t)=-erE(r_{particle})
        \end{equation}
        with the electric charge e, the position of the particle
        center at $r_{particle}$ and $r$ as the relative position
        of the charge from the particle center and the electric
        field vector $E$.
        For more general case, including large molecules the
        electric field can be better expressed with the vector
        potential $A(r,t)$ which is source free:
        \begin{equation}
            \label{eq17}
            \nabla\cdot A(r,t)=0
        \end{equation}
        and the electric field follows from this potential by:
        \begin{equation}
            \label{eq18}
            E(r,t)=-\frac{1}{c}.\frac{\partial}{\partial t}A(r,t)
        \end{equation}
        and the magnetic field by:
        \begin{equation}
            \label{eq19}
            H(r,t)=rot A(r,t)
        \end{equation}
        With respect to the quantum description, the vector
        potential can be written as:
        \begin{equation}
            \label{eq20}
            A(r,t)=\sum_{m}e_{m}\sqrt{\frac{h\lambda_{m}}{8\pi^2V\varepsilon_{0}c_{0}}}[b_{m}e^{\imath k_{m}r}+b_{m}^{+}e^{-\imath k_{m}r}]
        \end{equation}
        with the counter $m$ for the different waves of light and
        thus of the electric field, $e_{m}$ as the direction of
        the field vector, $\lambda_{m}$ as the wavelength of the
        light wave, V as the volume the waves are generated in and
        $k_{m}$ as the wave vector of the mth wave. The $b_{m}$
        and $b_{m}^{+}$ are photon absorbtion and emission
        operators which would be light amplitudes in the classical
        case. These operators fulfill the following relations:
        \begin{eqnarray}
            \label{eq21}
            b_{m}.b_{p}^{+}&-&b_{p}^{+}.b_{m}=\delta_{mp}
            \nonumber\\
            b_{m}.b_{p}&-&b_{p}.b_{m}=0
            \nonumber\\
            b_{m}^{+}.b_{p}^{+}&-&b_{m}^{+}.b_{m}^{+}=0
        \end{eqnarray}
        which result in the description of the energy of the
        electrical field by a sum over harmonic oscilators as:
        \begin{equation}
            \label{eq22}
            H_{field}=\sum_{m}b_{m}^{+}b_{m}h\nu_{m}
        \end{equation}
        and the Hamilton operator for a single electron in the
        potential of the cores V and the electric field $\bf{A}$ is
        given by:
        \begin{equation}
            \label{eq23}
            H_{electron}=\frac{1}{2m_{electron}}.[p-e_{e}A(r,t)]^2+V(r,t)
        \end{equation}
        with the mass $m_{electron}$ and charge $e_{e}$ of the
        electron and the pulse operator:
        \begin{equation}
            \label{eq24}
            p=-\frac{\imath}{\hbar}\nabla
        \end{equation}
        With these definitions the interaction operator for a
        one-electron system follows from:
        \begin{eqnarray}
            \label{eq25}
            H_{interaction}(r,t)&=&-\frac{e_{e}}{m_{electron}}A(r,t).p
            \nonumber\\
            &&{}+\frac{e_{e}^2}{2m_{electron}}A^{2}(r,t){}
        \end{eqnarray}
        for linear interactions the second term can be neglected.
        But the interaction has to be considered for all charges
        in the particle which are in molecules for all molecules
        and core charges.
        The resulting interaction operator is given by:
        \begin{eqnarray}
            \label{eq26}
            H_{interaction}(r,t)&=&\sum_{p}[-\frac{e_{e}}{m_{electron}}A(r_{p},t).p_{p}]
            \nonumber\\
            &=&+\sum_{q}[-\frac{Z_{core,q}e_{e}}{M_{core}}A(R_{q},t).P_{q}]
            \nonumber\\
       \end{eqnarray}
        with the charge $Z_{core,q}$ of the qth core, the
        coordinate $R_{q}$ of this core and its momentum $P_{q}$.
        In the dipole approximation the interaction operator for
        such a system can be written as:
        \begin{equation}
            \label{eq27}
            H_{interaction}(r,t)=\sum_{m}[-\frac{Z_{charge,m}e_{e}}{m_{charge,m}}E(r_{p},t)]
        \end{equation}
        and thus the transition dipole moment in the dipole
        approximation follows as:
        \begin{equation}
            \label{eq28}
            \mu_{p\leftarrow m}=e_{e}.\int_{V}\varphi_{p}^{*}.(\sum_{m}[-\frac{Z_{charge,m}e_{e}}{m_{charge,m}}E(r_{p},t)]).\varphi_{m}dV
        \end{equation}
        For a real material such as NPP or MNA this formula would be very complicated and take
        enormous calculations. Therefore some approximation must be
        applied. It can be shown that for absorbtion or emission of photons
        the material has to perform a transition between two
        eigenstates $E_{m}$ and $E_{p}$ of the material and thus
        the photon energy $E_{photon}$ has to fulfill the
        resonance condition:
        \begin{equation}
            \label{eq29}
            E_{photon}=h\nu_{photon}=|E_{p}-E_{m}|
        \end{equation}
        But for our linear phenomenon, the photon energy is about
        $2eV$ (in $\lambda=630nm$). If electron would be in HOMO (Highest Occupied
        Molecular Orbital), this electron do not go to LUMO (Lowest
        Unoccupied Molecular Orbital) or exited state by interaction.
        This phenomena is named nonresonant phenomenon,\cite{9,40,41} (nonresonant phenomena is
        not exclusive for nonlinear optical phenomena). Therefor
        electron after interaction, go to quasi states that their life times
        is very short, then  this electron go back to primary state after very short
        time. The nonresonant lifetime is determined by the
        uncertainty principle and the energy mismatch between
        photon energy in second time and the input photon energy.
        We can assume that the characteristic response time of
        this process is the time required for the electron cloud
        to  become distorted in response to an applied optical
        field. This response time can be estimated as the orbital
        period of the electron in its motion around the nucleus which is about $\tau\simeq10^{-16}s$ or
        100as,\cite{41}.
        We can estimate this characteristic response time according to (9) if, $n=3, L=3\mu$. Consequently
        $\Sigma\tau$ is equal to $10^{-14}$sec. Because in b direction of crystal in $3\mu$m length, approximately 4024
        molecules exist, therefore the average quantity of $\tau$:
        \begin{displaymath}
              \overline{\tau}=\frac{\sum\tau}{N}
        \end{displaymath}
        is in order of $10^{-18}s$ or 1 as. The perturbation in this very short time can assume semiclassically.
        In linear phenomenon in this short time, just one photon
        interacts with one molecule. Because NPP molecule has delocalization electrons,(or $\pi$-electron system),
        in benzene ring, that photon interacts with this electron type,\cite{32} and it is annihilated \cite{38}. We call this photon,
        a successful photon, (that does not produce phonon).\\
        To obtain $\pi$-electron wavefunction for benzene molecule the Schr\"{o}dinger equation may be solved.
        Since this is very complicated process, it cannot be done exactly, an approximated procedure known as H\"{u}ckle method
        must be employed. In this method, by using H\"{u}ckle Molecular-Orbital (HMO) calculation, a wave function
        is formulated that is a linear combination of the atomic orbitals (LCAO) that have overlapped \cite{37} (see
        Fig.3):
        \begin{equation}
            \label{eq31}
            \Psi=\sum_{i}C_{i}\Phi_{i}
        \end{equation}
        where the $\Phi_{i}$ refers to atomic orbitals of carbon atoms in the ring and the summation is over the six C
        atoms. The coefficients of the atomic orbitals are calculated
        self-consistently through Roothaan's equations to obtain the $\Psi$ and
        the corresponding the one electron energies $\epsilon$, \cite{43}:
        \begin{equation}
            \label{eq32}
            \sum_{\nu}(F_{\mu\nu}-\epsilon.S_{\mu\nu})C_{\nu}=0,
        \end{equation}
        where the Fock matrix $F_{\mu\nu}$ and overlap integrals
        $S_{\mu\nu}$ are given by:
        \begin{equation}
            \label{eq33}
            F_{\mu\nu}=H_{\mu\nu}+\sum_{\lambda\sigma}P_{\lambda\sigma}[(\mu\nu|\lambda\sigma)-\frac{1}{2}(\mu\lambda|\nu\sigma)]
        \end{equation}
        and
        \begin{equation}
            \label{eq34}
            S_{\mu\nu}=\int\phi_{\mu}.\phi_{\nu}d\tau
        \end{equation}
        Here the core Hamiltonian matrix $H_{\mu\nu}$, density
        matrix $P_{\mu\nu}$, and two-electron repulsion integrals
        are given by:
        \begin{equation}
            \label{eq35}
            H_{\mu\nu}=\int\phi_{\mu}H^{core}\phi_{\nu}d\tau,
        \end{equation}
        \begin{equation}
            \label{eq36}
            P_{\mu\nu}=2\sum_{i=1}^{occupied}C_{\mu i}C_{\nu i},
        \end{equation}
        and
        \begin{equation}
            \label{eq37}
            (\mu\nu|\lambda\sigma)=\int\phi_{\mu}(1)\phi_{\nu}(1)\frac{1}{r_{12}}\phi_{\lambda}(2)\phi_{\sigma}(2)d\tau_{1}d\tau_{2},
        \end{equation}
        and the molecular Hamiltonian by
        \begin{equation}
            \label{eq38}
            H=\sum_{i}H_{i}^{core}+\sum_{i<j}\frac{1}{r_{ij}},\
            H_{i}^{core}=-\frac{1}{2}\nabla_{i}^{2}-\sum_{A}\frac{Z_{A}}{r_{Ai}}
        \end{equation}
        where the sum on $i(A)$ is over electrons (nucleii), and
        the $Z_{A}$ is the net core charge. The $|C_{i}|^{2}$ is the probability of the $\pi$-electron at $i$th
        atom. Thus:
        \begin{displaymath}
        \label{eq11}
            |C_{1}|^{2}+|C_{2}|^{2}+|C_{3}|^{2}+|C_{4}|^{2}+|C_{5}|^{2}+|C_{6}|^{2}=1
        \end{displaymath}
        In the case of Benzene molecule:
        \begin{displaymath}
          |C_{i}|^{2} =\frac{1}{6}
        \end{displaymath}
        as followed from the symmetry of the ring \cite{44,45}.
        But NPP and MNA molecules aren't such as Benzene molecule. NPP is polar molecule. Nitro $(NO_{2})$ is more
        powerful electronegative compound than prolinol and pulls $\pi$-electron system; consequently, the probability
        of finding $\pi$-electron system at various carbon atoms of main ring isn't the same and the probability of finding
        $\pi$-electrons near the Nitro group is greater than near the prolinol group. Therefore there is no symmetry
        for NPP and electron cloud is spindly or oblong, (similar to dom-bell) (Fig.\ref{Fig6}).
        \begin{figure}
            \centering
            \includegraphics[scale=0.9]{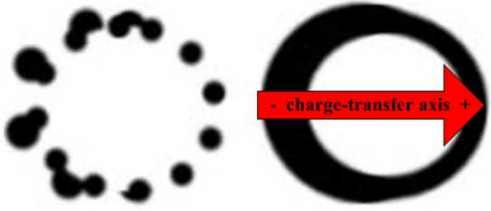}
            \caption{\footnotesize Assumed $\pi$-electron orbit of NPP molecule that obtained from fig. 3 for Benzene molecule
            (approximately).}\label{Fig6}
        \end{figure}
        We estimate this form of electron cloud by an ellipse
        that our calculations would be uncomplicated. We assume effective positive
        charge that is located in one of focal points of ellipse.
        The quantity of this effective positive charge is
        determined by semiclassical arguments.
        For attaining probability of electron presence on an orbit (Fig.\ref{Fig7}),
        \begin{figure}
            \centering
            \includegraphics[scale=0.7]{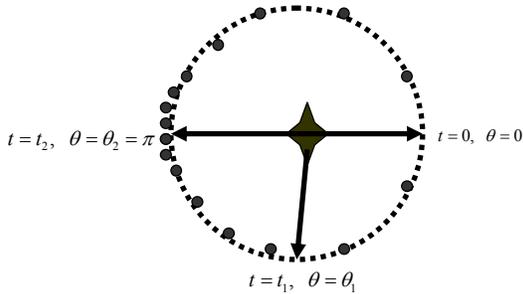}
            \caption{\footnotesize $\pi$-electron system approximation for NPP molecule, $\theta$ = 0 is in the positive
            direction of dipole (Prolinol side) and $\theta$ = $\pi$ is in the negative
            direction of dipole (Nitro side).}\label{Fig7}
        \end{figure}
        we say, T time is required by radial vector to sweep total $\pi.u.v$
        interior area of ellipse (u and v are semimajor and semiminor axis of ellipse respectively),
        in t times, this radial vector sweeps:
        \begin{displaymath}
          \pi.u.v.\frac{t}{T}
        \end{displaymath}
        area of ellipse, (see Fig.\ref{Fig7}). If t is the time, that electron sweeps $\theta$ radian of orbit
        then t is obtained from this relation \cite{32}:
        \begin{equation}
            \label{eq39}
            t=\frac{T}{2\pi}\{2\arctan(\sqrt{\frac{1-\varepsilon}{1+\varepsilon}}\tan(\frac{\theta}{2}))-
            \frac{\varepsilon.\sqrt{1-\varepsilon^{2}}.\sin(\theta)}{1+\varepsilon.\cos(\theta)}\}
        \end{equation}
        Where $\varepsilon$ is ellipse eccentricity. By using this relation, we attain the required
        time (t) for electron to traverse from $\theta$ to $\theta+d\theta$ and it is divided
        by total time T. By this approach, we can determine the PDF (Probability Density Function) approximately. The PDF in apogee
        (near the Nitro group), is maximum and in perigee (near the Prolinol or Methane) is
        minimum. Therefore PDF is correlated to $\theta$ from (38) and seen in Fig.\ref{Fig8}.
        \begin{figure}
            \centering
            \includegraphics[scale=0.31,width=3.1in,height=2.2in]{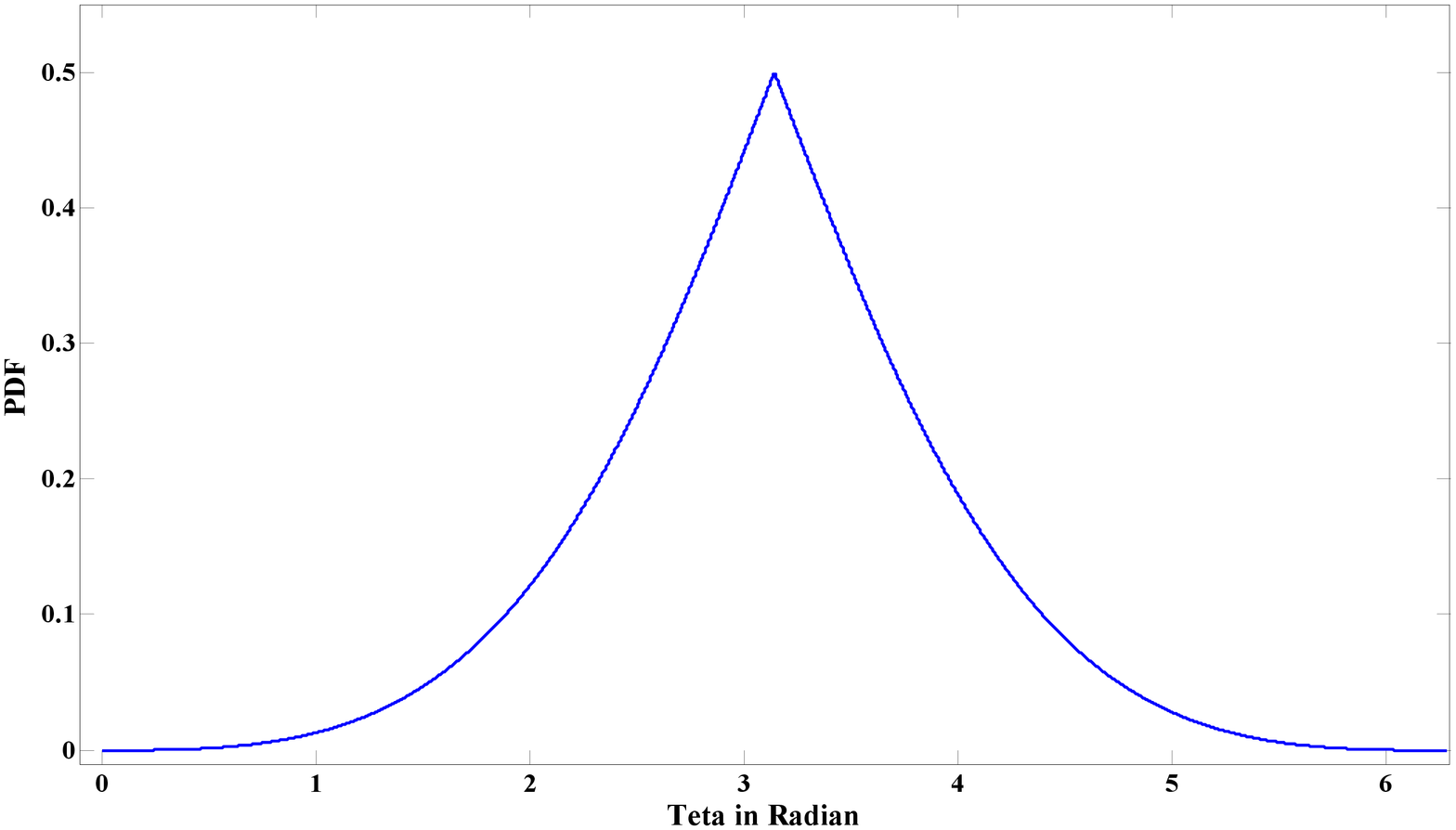}
            \caption{\footnotesize PDF of presence probability of $\pi$-electron in different energy states when it rotates around the
            assumed orbit in Fig.\ref{Fig7}; as we see the PDF is correlated to $\theta$. }\label{Fig8}
        \end{figure}
        The $\tau$ quantity is correlated to $\theta$ and consequently, $\tau$
        quantity is correlated to presence probability of
        $\pi$-electron system.\\
        The angle between Y vector and charge transfer action (N$_{1}$-N$_{2}$) is 58.6$^\circ$ and X and Z
        axis is perpendicular to Y (Fig.\ref{Fig9}).
        \begin{figure}
            \centering
            \includegraphics[scale=0.4]{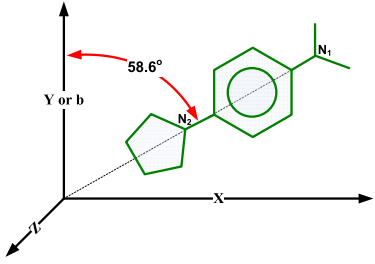}
            \caption{\footnotesize X, Y and Z axis and NPP molecule in dielectric frame.}\label{Fig9}
        \end{figure}
        We consider propagation along Z direction. We spot a photon interacts
        with $\pi$-electron of NPP in first layer. After interaction, this photon gives
        its energy to electron and is annihilated. Electron absorbs energy and digresses in direction of photon momentum.
        Electron with photon energy, may not be unbounded and after arriving to apogee of digression, it returns back to
        ground state, because the photon energy is equal to $h\nu=1.96ev$ (h is Planck's constant and $\nu$ is frequency of
        laser beam) whereas energy for excitation is greater than $3ev$. When electron returns to
        ground state one photon is produced. The time coming up and down is $\tau$ delay time. This photon after freedom goes to second
        layer in direction of annihilated photon (nonce, we assume the polarization doesn't change), in second layer
        this photon interacts with another delocalization $\pi$-electron
        certainly, because the effective interaction range of photon is approximately equal to its wavelength and is very greater than
        the distance between molecules. This molecule is nearest to photon effective central.
        This action is repeated for each layer. The location of photon-electron
        interaction is significant in every molecule and it is effective on $\tau$ quantity directly.
        We assume that interacting photon has circle polarization and electron subject to virtual positive charge
        center. The phase retardation between $E_{x}$ and $E_{y}$ (9)
        can be obtain from this relation,\cite{32}:
        \begin{equation}
            \label{eq41}
            \Delta\phi=\frac{\omega.\sqrt{2h\nu.m}}{KZe^{2}}\sum_{i=1}^{m}[\cos(\theta_{i})-\sin(\theta_{i})].r_{i}^{2}
        \end{equation}
        that
        \begin{displaymath}
            r_{i}=\frac{(1-\epsilon^{2}).u}{1+\epsilon.\cos(\theta_{i})}
        \end{displaymath}
        where $\epsilon$ is the elliptical eccentricity
        and u is the semimajor axis of the ellipse. By applying an
        external transverse electric field to organic crystal (in the range of several volts per
        micron) the shape of $\pi$-electron system will be
        deformed slightly and we would be expect some noticeable
        variations in microscopic delay parameters ($\tau_{x}$,$\tau_{y}$)
        and phase retardation; (see Fig.10).
        \begin{figure*}
            \centering
            \includegraphics[scale=0.4]{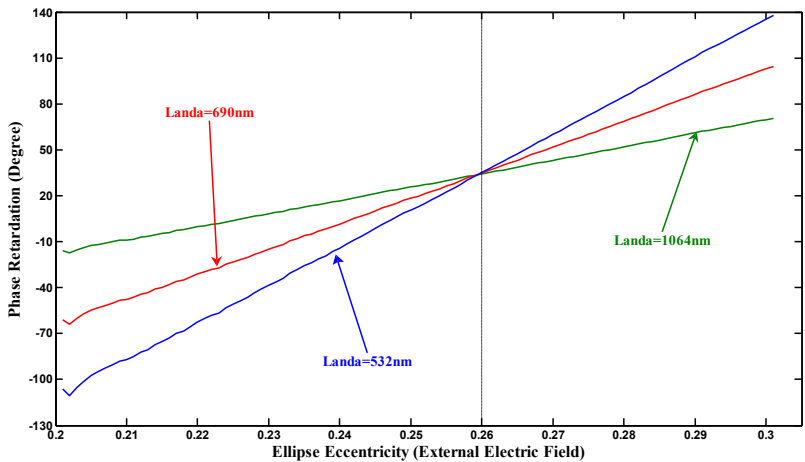}
            \caption{\footnotesize Phase retardation between $E_{x}$ and $E_{y}$ of optical signal with different
            wavelength by applied external electric field. As we
            see; by applied electric field, the virtual ellipse eccentricity change (electron in electrical field). With eccentricity
            variation, the microscopic delay ($\tau_{x}$,$\tau_{y}$) change and consequently phase retardation is modified.}\label{Fig10}
        \end{figure*}
        Consequently, by step-like change of input
        voltage, optical signal is switched between output port of
        DOS.

        \section{\large Variation Analysis of Phase Retardation versus Applied Electric Field}
        We simulate phase retardation of $3\mu m$-length NPP crystal by Monte-Carlo method, then we generate random
        number using $MATLAB$ program. This program produces PDF quantities was explained in before subsection
        and relates each of them to every molecule. These values are indexing $\pi$-electron positions in each
        layer, by assumption a reference point (see fig. 6). Additionally we have used a $MATLAB$ program for Monte-Carlo
        simulation. The inputs of this program are:\\ \indent
        1.  The wavelength of incident optical beam in which we want to design DOSs;\\ \indent
        2.  $h, m, q, k=\frac{1}{4\pi.\varepsilon}, c_{\scriptscriptstyle{0}}$ that are Planck's constant,
        electron rest mass, elementary charge, Coulomb constant and speed of light respectively.\\ \indent
        3.  Unit cell parameters of NPP crystal: a, b, c, $\beta$ and its other parameters that have given in
        subsection (A).\\ \indent
        4.  L: crystal thickness that in our simulation it is 3$\mu$m.\\
        And the outputs of
        $MATLAB$ program are: phase retardation in each wavelength.\\ \indent
        System calibration is done semiclassically by experimental refractive index
        data.t
        In this method that we obtain three refractive indexes with x-polarization in
        threea
        with $\epsilon$ (eccentricity), u (semimajor axis of ellipse) and Z (equivalent positive charge) in a
        way that refractive indexes in three wavelengths are very close to experimental data.
        Then we would see that refractive index in other
        wavelengths and other polarization with same $\epsilon$, u and Z will be achieved. Of course these values, $\epsilon$, u and
        Z would be close to experimental structure of crystal, for example u would be greater than and smaller than
        minimum and maximum sizes of six lengths of benzene hexagonal respectively, or $\epsilon$ would be small
        but greater than zero. In other hand these values must be logical. From this method in our simulation we have
        obtained $\epsilon=0.26, Z=3.9, u=1.4A^{0}$ that is very close to experimental and structural
        data.\\ \indent
        Xu and co-workers \cite{19,26}, have done some electro-optic experiment about single crystal film
        of NPP. They have obtained
        $|n_{x}^{3}r_{12}-n_{y}^{3}r_{22}|=340 pm/V$ and
        $r_{12}=65 pm/V$ in an optical beam with 1064nm wavelength. They have studied phase retardation
        between $E_{x}$ and $E_{y}$ of optical beam as a function
        of angle between electric field and charge transfer action
        of NPP. They have concluded that the maximum phase retardation was observed for the field oriented
        along the charge-transfer axis which was parallel to the film surface. The electro-optic effect
        or phase retardation was negligible when the electric field was
        applied perpendicular to the charge-transfer axis.
        This concept could be justified by our model in previous
        subsection. When the angle between charge-transfer action
        and external electric field change, the ellipse
        eccentricity modify and consequently, the phase
        retardation alter. Obviously, from Fig.5 when the external electric
        is parallel to charge-transfer axis, the ellipse drag more
        and the ellipse convert to a line. Therefor, ellipse
        eccentricity arise and from Fig.9 the phase retardation
        growth. In the other hand, from Fig.5 when the angle
        between external electric field and charge-transfer axis
        change, the ellipse is gathered and convert to circle and the eccentricity
        decrease to zero. Thus from Fig.9 the phase retardation is
        lowered.

       \section{\large Conclusion}

        we justified linear EO phenomenon by QPM. This suggested physical model could be a
        powerful tool for analyzing and explaining processes that happen in waveguides with microscopic and nanoscopic sizes.
        We showed how the phase retardation between different arguments of an optical field with distinctive wavelengths can take place.

 \end{document}